# Free-breathing 3D cardiac extracellular volume (ECV) mapping using a linear tangent space alignment (LTSA) model

Wonil Lee[1†], Paul Kyu Han[1†], Thibault Marin[1], Ismaël B.G. Mounime[1,2], Samira Vafay Eslahi[3,4], Yanis Djebra[1], Didi Chi[1], Felicitas J. Bijari[1]

Marc D. Normandin[1], Georges El Fakhri[1], Chao Ma[1*]

[1] Department of Radiology and Biomedical Imaging, Yale School of Medicine, Yale University, New Haven, USA

[2] LTCI, Télécom Paris, Institut Polytechnique de Paris, France

[3] Gordon Center for Medical Imaging, Department of Radiology, Massachusetts General Hospital, Boston, Massachusetts, USA

[4] Department of Radiology, Harvard Medical School, Boston, Massachusetts, USA

†These authors contributed equally to this work.

*Correspondence to:    Chao Ma, PhD

Department of Radiology and Biomedical Imaging

Yale School of Medicine

100 Church Street South

New Haven, CT, 06519

E-mail: chao.ma.cm2943@yale.edu

Word count of the manuscript body: 4496 words

# Abstract

**Purpose:** To develop a new method for free-breathing 3D extracellular volume (ECV) mapping of the whole heart at 3T.

**Methods:** A free-breathing 3D cardiac ECV mapping method was developed at 3T. T1 mapping was performed before and after contrast agent injection using a free-breathing ECG-gated inversion-recovery sequence with spoiled gradient echo readout. A linear tangent space alignment (LTSA) model-based method was used to reconstruct high-frame-rate dynamic images from (k,t)-space data sparsely sampled along a random stack-of-stars trajectory. Joint T1 and transmit B1 estimation was performed voxel-by-voxel for pre- and post-contrast T1 mapping. To account for the time-varying T1 after contrast agent injection, a linearly time-varying T1 model was introduced for post-contrast T1 mapping. ECV maps were generated by aligning pre- and post-contrast T1 maps through affine transformation.

**Results:** The feasibility of the proposed method was demonstrated using in vivo studies with six healthy volunteers at 3T. We obtained 3D ECV maps at a spatial resolution of $1.9 \times 1.9 \times 4.5$ mm$^3$ and a FOV of $308 \times 308 \times 144$ mm$^3$, with a scan time of $10.1 \pm 1.4$ and $10.6 \pm 1.6$ min before and after contrast agent injection, respectively. The ECV maps and the pre- and post-contrast T1 maps obtained by the proposed method were in good agreement with the 2D MOLLI method both qualitatively and quantitatively.

**Conclusion:** The proposed method allows for free-breathing 3D ECV mapping of the whole heart within a practically feasible imaging time. The estimated ECV values from the proposed method were comparable to those from the existing method.

# 1. Introduction

Cardiac T1 mapping provides critical information about the characteristics of the myocardium. The pre-contrast T1 value is a valuable biomarker for diagnosing cardiac diseases characterized by changes in tissue composition, such as cardiac hemochromatosis, amyloidosis, myocarditis, and Anderson-Fabry disease (1-3). Extracellular volume (ECV) fraction, which can be measured using the T1 before and after contrast agent injection, can provide additional diagnostic information associated with the myocardial interstitium. ECV can be useful for detecting diseases such as myocardial fibrosis, which can be challenging to detect using the conventional qualitative late gadolinium enhancement (LGE) method (2,3). Furthermore, knowing the ECV information can be beneficial in monitoring disease progression (4) or prognosis prediction (5).

Despite its usefulness, cardiac T1 mapping is challenging due to cardiac and respiratory motion. To date, extensive research efforts have been made to overcome this issue. Several approaches have been proposed to estimate T1 values capturing saturation recovery and inversion recovery signals across multiple heartbeats with ECG-gating (6-11). Modified look-locker inversion recovery (MOLLI) is a widely accepted technique that utilizes ECG-gating and breath-holding to allow cardiac T1 mapping in 2D (6). Efforts have been made to improve the spatial coverage of this method with the help of simultaneous multi-slice imaging, compressed sensing, and parallel imaging techniques (12-14). Methods with ECG-gating and navigators to handle respiratory motion have been developed to allow cardiac T1 mapping with free-breathing acquisition for improved subject comfort (15-25). Look-locker-based methods with continuous data acquisition have been developed to reduce imaging time (26-32). Recently, deep learning has been employed for cardiac T1 mapping to reduce the scan time and estimation time of T1 (33-43).

Although various methods have been proposed for 3D pre-contrast cardiac T1 mapping, relatively fewer efforts have been made to enable 3D post-contrast T1 or ECV mapping of the heart. To date, post-contrast cardiac T1 mapping methods have been limited to 2D (44,45) or 3D imaging with limited spatial coverage (15,23), and ECV mapping methods have been limited to 2D imaging (44,45). The major challenge is the variation of T1 after administration of the contrast agent. When the contrast agent is administered with the typical bolus injection, post-contrast cardiac T1 and ECV mapping are recommended to be performed with sufficient delay (e.g., at least 10 minutes) after the injection (2) and short imaging time (15,23) since the T1 of the myocardium and blood can vary significantly after injection (e.g., several tens of milliseconds per minute) (45). The 3D cardiac T1 mapping methods developed for the case without the involvement of a contrast agent typically have a relatively long acquisition time (e.g., several to tens of minutes) and, therefore, cannot be directly applied for 3D post-contrast T1 and ECV mapping. To our knowledge, no methods have been proposed to accommodate the T1 changes over time after contrast agent injection and allow post-contrast T1 and ECV mapping of the heart in 3D.

This work presents a novel method to enable 3D free-breathing cardiac ECV mapping. Cardiac T1 mapping was performed using a free-breathing ECG-gated inversion-recovery sequence with spoiled gradient echo readout before and after administering contrast agent. The (k,t)-space was sparsely sampled using a random stack-of-stars trajectory to reduce imaging time. Image reconstruction was performed using a linear tangent space alignment (LTSA) model-based framework, which leverages the intrinsic low-dimensional manifold structure of the underlying signals to recover real-time dynamic MR images with high framerate from highly undersampled data. After image reconstruction, joint T1 and transmit B1 (B1+) estimation was performed voxel-by-voxel for pre- and post-contrast T1 mapping. To account for the time-varying T1 after contrast agent injection, a linearly time-varying T1 model was introduced for post-contrast T1 mapping. The feasibility of the proposed method was demonstrated using *in vivo* studies with six healthy volunteers at 3T.

# 2. Methods

## 2.1 Data acquisition

All experiments were performed under a study protocol approved by the local institutional review board (IRB). Six healthy volunteers (1M and 5F; 38 ± 15 years) were imaged using a 3T MRI scanner (Biograph mMR, Siemens Healthcare, Erlangen, Germany). Blood sampling was performed prior to imaging to perform pregnancy test and/or to measure the hematocrit level for each subject. Cardiac T1 mapping was performed using a free-breathing, ECG-gated, inversion recovery (IR) sequence with a 3D spoiled gradient-echo (SPGR) readout and sparse-sampling (21) as shown in Figure 1, before and after the contrast agent injection (Dotarem®, 0.1 mmol/kg), to measure ECV. The imaging parameters were as follows: field-of-view (FOV) = 308×308×144 mm$^3$, spatial resolution = 1.9×1.9×4.5 mm$^3$, flip angle (FA) = 9°, TR = 4.2 ms, TE = 1.7 ms, 10-(3)-10-(3) protocol, and acquisition window per frame = 138.6 ms. Two frames were acquired per cardiac cycle at end-diastolic phase to maximize data acquisition efficiency, with training and imaging datasets acquired in an interleaved fashion over 1 and 32 TRs for each frame, respectively, as shown in Figures 1b and 1c. The imaging dataset collected 32 spokes over random k$_z$ locations and in-plane spoke angles between $0°$ and $180°$ in each frame (corresponding to an effective acceleration factor of 251 for each frame) (Figure 1c). Post-injection scans were performed 8.8, 17.2, 19.3, 16.8, 19.0, and 19.2 min after the contrast agent injection (corresponding to the midpoint of the data acquisition between 10 to 30 minutes after the contrast agent injection (46), which was 13.8, 21.9, 24.7, 22.4, 25.1, and 23.3 minutes after the contrast agent injection) for Subjects 1 to 6, respectively. The different delays among the subjects were due to variations in the subject heart rate and practical imaging conditions during the experiments. The data from Subject 1 was acquired at a relatively earlier time than the others because the study was part of a simultaneous PET/MR study that had additional timing constraints, e.g., PET tracer injection and blood sampling. Additional imaging was performed using the MOLLI protocol in (47) before and/or after the acquisition of the proposed method for pre- and post-contrast injection cases to get 2D T1 maps for comparison. The whole imaging protocol is shown in Figure 1, where the timing of each cardiac T1 map is labeled for the convenience of future references.

## 2.2 LTSA-based image reconstruction

We assume that the temporal signal of a given voxel in dynamic MR images lies on a $D$-dimensional manifold $\mathcal{F}$ ($D \ll N$, with $N$ the total number of frames acquired) (48,49):

$$\boldsymbol{x}_m = f(\boldsymbol{z}_m), \tag{1}$$

where $\boldsymbol{x}_m = [x(\boldsymbol{r}_m, t_1) \dots x(\boldsymbol{r}_m, t_N)]^\mathrm{T} \in \mathbb{C}^{N \times 1}$ denotes the temporal signal at position $\boldsymbol{r}_m$ ($m = 1, \cdots, M$, with $M$ the total number of voxels ), $\boldsymbol{z}_m \in \mathbb{C}^{D \times 1}$ the corresponding global coordinates in the feature space, and $f$ the unknown non-linear mapping between the signal space and the feature space. Assuming smoothness and regularity (i.e., $f$ is differentiable and its Jacobian matrix is of full rank $D$), $\mathcal{F}$ can locally be approximated using the first-order Taylor expansion in the feature space:

$$\boldsymbol{x}_m^{(q)} = \boldsymbol{\Pi}_q f(\boldsymbol{z}_m) \approx \boldsymbol{\Pi}_q f(\overline{\boldsymbol{z}}_q) + \boldsymbol{\Pi}_q \boldsymbol{J}_f(\overline{\boldsymbol{z}}_q)(\boldsymbol{z}_m - \overline{\boldsymbol{z}}_q), \tag{2}$$

where $\boldsymbol{\Pi}_q \in \mathbb{R}^{P_q \times N}$ denotes an operator selecting a total of $P_q$ frames to form the $q$-th neighborhood, $\boldsymbol{x}_m^{(q)} = [x(\boldsymbol{r}_m, t_1^{(q)}), \dots, x(\boldsymbol{r}_m, t_{P_q}^{(q)})]^\mathrm{T} \in \mathbb{C}^{P_q \times 1}$ the temporal signals of voxel $m$ in the $q$-th neighborhood, $\{t_p^{(q)}\}_{p=1}^{P_q} \subset \{t_1, \dots, t_N\}$ the corresponding frame indices, $\overline{\boldsymbol{z}}_q$ the global coordinate of the centroid of the signals $\boldsymbol{x}_m^{(q)}$ in the $q$-th neighborhood (i.e., $\boldsymbol{\Pi}_q f(\overline{\boldsymbol{z}}_q) = \sum_{m=1}^{M} \boldsymbol{x}_m^{(q)}/M$), and $\boldsymbol{J}_f(\overline{\boldsymbol{z}}_q) \in \mathbb{C}^{N \times D}$ is the Jacobian of the nonlinear function $f$ evaluated at $\overline{\boldsymbol{z}}_q$. Note that the columns of $\boldsymbol{J}_f(\overline{\boldsymbol{z}}_q)$ span the tangent space of the manifold at $\overline{\boldsymbol{z}}_q$.

Equation (2) represents a local linear approximation of the temporal signals in the $q$-th neighborhood, which can be written in the following matrix/vector form:

$$\mathbf{X}_q \approx \mathbf{T} \mathbf{J}_q^{\mathrm{T}} \boldsymbol{\Pi}_q^{\mathrm{T}}, \tag{3}$$

where $\mathbf{T} \in \mathbb{C}^{M \times (D+1)}$ concatenates the augmented global coordinates of the temporal signals of all the voxels into a matrix and $\mathbf{J}_q^{\boxed{}} \in \mathbb{C}^{N \times (D+1)}$ denotes the augmented Jacobian matrix evaluated at $\overline{\boldsymbol{z}}_q$. $\mathbf{X}_q \in \mathbb{C}^{M \times P_q}$ is a Casorati matrix formed by the signals in the $q$-th neighborhood:

$$\mathbf{X}_q = \mathbf{X}\boldsymbol{\Pi}_q^{\mathrm{T}} = \begin{bmatrix} x(\boldsymbol{r}_1, t_1^{(q)}) & \cdots & x(\boldsymbol{r}_1, t_{P_q}^{(q)}) \\ \vdots & \ddots & \vdots \\ x(\boldsymbol{r}_M, t_1^{(q)}) & \cdots & x(\boldsymbol{r}_M, t_{P_q}^{(q)}) \end{bmatrix}, \tag{4}$$

where $\mathbf{X} \in \mathbb{C}^{M \times N}$ is the Casorati matrix formed by all the acquired data.

However, using Eq. (3) directly is intractable because the non-linear mapping $f$ between the signal space and feature space is implicitly defined and prevents Jacobian calculation. To address this issue, we leverage the low-rank approximation of the local Casorati matrix in Eq. (4):

$$\mathbf{X}_q \approx \boldsymbol{\Theta}_q \boldsymbol{\Phi}_q^{\mathrm{T}}, \tag{5}$$

where the columns of $\boldsymbol{\Phi}_q \in \mathbb{C}^{P_q \times (D+1)}$ contain the temporal basis functions spanning a subspace that approximates the tangent space of the manifold in the $q$-th neighborhood and $\boldsymbol{\Theta}_q \in \mathbb{C}^{M \times (D+1)}$ concatenates the corresponding local coordinates.

Combining Eqs. (3) and (5) leads to the following Linear Tangent Space Alignment (LTSA) model:

$$\mathbf{X} = \sum_{q=1}^{Q} \mathbf{X}_q \boldsymbol{\Pi}_q = \sum_{q=1}^{Q} \mathbf{T} \mathbf{L}_q \boldsymbol{\Phi}_q^{T} \boldsymbol{\Pi}_q, \tag{6}$$

where $L_q \in \mathbb{C}^{(D+1)\times(D+1)}$ denotes a matrix that aligns local coordinates $\Theta_q$ in the signal space to the global coordinates $T$ in the feature space.

Given the acquisition scheme described in Section 2.1, the temporal bases $\Phi_q$ can be predetermined by performing singular value decomposition (SVD) on the measured training data. The image reconstruction problem can be then formulated as the following constrained least-square optimization problem:

$$\arg\min_{T,L} \frac{1}{2} \left\| \Omega \left( F_s \sum_{q=1}^{Q} T L_q \Phi_q^T \Pi_q \right) - d \right\|_2^2 + \frac{\mu_T}{2} \| T \|_F^2 + \frac{\mu_L}{2} \| L \|_F^2$$

$$+ \lambda_T \| \mathcal{D}(TL) \|_1 + \lambda_L \| vec(L) \|_1$$

(7)

where $L = [L_1 \ldots L_Q] \in \mathbb{C}^{(D+1)\times Q(D+1)}$ concatenates all the alignment matrices, $d$ denotes the measured data in the $(k, t)$-space, $\Omega$ a $(k, t)$-space sampling operator, $F_s$ the Non-Uniform Fast Fourier Transform (NUFFT) operator (50), $\|\cdot\|_F$ the Frobenius norm, and $\mathcal{D}$ a finite difference operator. The objective function in Eq. (7) consists of data fidelity and regularization terms. Both $l_2$-norm terms are for numerical stability. A total variation penalty is imposed on the local coordinates (i.e., $TL$) to promote smoothness in the reconstructed images. In this work, the original LTSA (49) model is extended with an additional sparsity constraint on the $L$ matrix, leading to improved reconstructions by better selecting the tangent spaces to be aligned. A nested variation of the Alternating Direction Method of Multipliers (ADMM) (51) is used to solve the non-convex optimization in Eq. (7). More details on the algorithm can be found in Appendix.

In this work, the dynamic frames were grouped into different neighborhoods according to the respiratory movement, using the position of the tip of the liver as a surrogate signal. No overlapping blocks were used in this work. The dimensionality of the manifold was chosen empirically based on a trade-off between image quality and rank reduction. It is important to note that with the additional sparsity constraint, the rank $D$ becomes an upper bound of the neighborhood's intrinsic rank as it promotes sparse alignment matrix. The regularization parameters were selected using a grid search method. The combination of parameters offering the clearest delineation of the myocardium wall and the least amount of artifacts in the background was selected. The following regularization parameters were used to reconstruct dynamic images for all the subjects: $\mu_T = 1e^{-5}, \mu_L = 1e^{-10}, \lambda_T = 1e^{-10}, \lambda_L = 1e^{-12}$.

In this work, the reconstruction was executed on graphics processing units (GPUs) utilizing Python along with the CuPy library, incorporating customized kernels developed using the Computed Unified Device Architecture (CUDA). The solver utilized the SigPy framework and the NUFFT on GPUs was conducted using the MRRT package. The reconstructions were performed on a computing cluster containing four Tesla V100-SMX2 GPUs, each having 16GB of memory. These reconstructions were parallelized coil-by-coil across the four GPUs and subsequently integrated by computing the root-mean-square average of per-coil reconstructions.

## 2.3 Estimation of T1 and ECV

T1 estimation was performed voxel by voxel using an extension of the procedure described in (21). The number of time-point images or dynamic frames obtained after image reconstruction was 900, 900, 900, 900, 880, and 900, respectively, for Subjects 1 to 6 (which corresponds to acquisition times of 10.1$\pm$1.4 and 10.6$\pm$1.6 min for pre-and post-injection acquisitions, respectively). There were slightly fewer images for Subject 5 due to time constraints during the scan. The reconstructed dynamic images were in the same order as they were acquired. Reconstructed images were then binned into eight respiratory phases, according to the tip of the liver position estimated from the navigator signal described in Figure 1a. For pre-contrast acquisitions, there were 102$\pm$10, 116$\pm$7, 114$\pm$6, 110$\pm$4, 112$\pm$5, 114$\pm$5, 112$\pm$6, and 114$\pm$8 time-point images in each respiratory phase across the 6 subjects after binning. For post-contrast acquisitions, there were 108$\pm$4, 109$\pm$10, 112$\pm$6, 109$\pm$6, 113$\pm$5, 112$\pm$6, 113$\pm$6, and 114$\pm$15 time-point images for each respiratory phase across the 6 subjects after binning. T$_1$ mapping was performed for each respiratory phase via matching the signal from the reconstructed images with dictionary bases simulated by solving the Bloch equation numerically using the actual timing of data acquisition recorded by the scanner during the experiment. For post-contrast T1 estimation, we assumed T1 varies linearly with time. Joint estimation of T1, flip angle (FA), and rate of T1 change ($s$) was performed by fitting the reconstructed dynamic signals using the variable projection method (52), which led to the following optimization problem:

$$\{\widehat{T}_{1,m,b}, \widehat{FA}_{m,b}, \hat{s}_{m,b}\} = \arg\max_{T_1, FA, s} \frac{\left|x_{m,b}^{\mathrm{H}} \boldsymbol{\alpha}(T_1, FA, s)\right|}{\|\boldsymbol{\alpha}(T_1, FA, s)\|_2^2}, \tag{8}$$

where $x_{m,b}$ denotes the dynamic signal at pixel $m$ and bin $b$ and $\boldsymbol{\alpha}(T_1, FA, s)$ denotes a dictionary of basis functions synthesized from Bloch equation simulations for a grid of $(T_1, FA, s)$ triplets using the time stamps recorded during the data acquisition. The basis functions were synthesized using a grid of T1 with a stepsize of 10 ms from 500 ms to 2500 ms for the pre-contrast case and from 200 ms to 1500 ms for the post-contrast case, $FA = B1 \times FA_0$ with B1 from 0.2 to 1.5 with a stepsize of 0.05, and $s$ from 0 ms/s to 1 ms/s with a stepsize of $5 \times 10^{-2}$ ms/s. For pre-contrast T1 mapping, $s$ was set to 0, since T1 values are constant over time in native T1 mapping. Using a workstation computer with Inter (R) Xeon (R) Platinum 8268 CPU 2.8 GHz, the time for dictionary generation and fitting was $1.00 \pm 0.31$ seconds and $1413.60 \pm 59.77$ seconds, respectively, for pre-contrast case. The time for dictionary generation and fitting was $11.06 \pm 3.62$ seconds and $2780.90 \pm 384.4$ seconds, respectively, for post-contrast case. Prior to ECV calculation, image registration was performed using affine transformation to register the pre- and post-contrast T1 maps. ECV was then calculated for each voxel using the pre- and post-contrast T1s from each voxel, pre- and post-contrast T1s from the blood, and the measured hematocrit level for each subject using the following equation (47):

$$ECV(\%) = (1 - HCT)\frac{\frac{1}{T1_{myo,post}} - \frac{1}{T1_{myo,pre}}}{\frac{1}{T1_{blood,post}} - \frac{1}{T1_{blood,pre}}} \times 100, \tag{9}$$

## *2.4 Analysis*

All analyses were performed using all of the frames in each respiratory phase. Sixteen region-of-interests (ROIs) were drawn in the myocardium for each subject according to the AHA recommendations (53), excluding the apex due to difficulties in accurate comparison caused by banding artifacts observed in MOLLI at the apex. The mean T1 was calculated for each ROI for assessment.

# 3. Results

Figures 2 and 3 show the representative results of image reconstruction using the low-rank (LR) based method with sparsity constraints and the proposed LTSA method. Overall, images were successfully reconstructed from under-sampled (k,t)-space data using both methods without any significant artifacts in the images across different frames (Figure 2). The images reconstructed from LTSA showed improvement over those from LR and LTSA without the sparsity constraint, displaying clearer delineation of the myocardium wall and less artifacts in the chest (Figure 2, Supporting Information Figure S1). These improvements were also reflected in the estimated T1 and ECV maps, where the results obtained by the LTSA method showed better agreement with those from the MOLLI method while the LR method overestimated pre-contrast T1 values in both the myocardium and blood pool and ECV values in the blood pool (Supporting Information Figures S2 and S3). Similar image reconstruction quality was observed using LTSA for pre- and post-contrast datasets across different frames and slice locations, as shown in Figure 3.

Figures 4 to 7 show pre- and post-contrast T1 maps as well as ECV maps from Subject 3 obtained by the proposed method. The pre-contrast T1 map was successfully acquired in 3D using the proposed method, with visually similar estimation of myocardial and blood T1 compared to those from MOLLI (Figure 4). Similarly, the post-contrast T1 map was successfully estimated in 3D using the proposed method (Figure 5). The estimated T1 values from the proposed method were slightly lower than those from MOLLI for the post-contrast case, presumably due to the difference in acquisition time post-contrast agent injection (Figure 5). Figure 6 shows the results of T1 estimation over time after the administration of the contrast agent. The proposed method estimated an increase in myocardial T1 over time post-contrast agent injection (Figure 6a). These values were within a comparable range with respect to the estimated myocardial T1 from MOLLI acquired at times before and after the proposed method (Figure 6a). The same findings were observed qualitatively over different tissue types, as shown from the estimated T1 maps (Figure 6b). Figure 7 shows the results of ECV estimation using the proposed method. The ECV map was successfully estimated in 3D using the proposed method, covering the basal end to the apical end of the heart (Figure 7). The ECV maps generated from the proposed method were visually comparable to those from MOLLI at the same slice locations (Figure 7). Similar findings were observed from additional two subjects (Figure 8).

Figures 9 and 10 show a quantitative comparison of the estimated pre-contrast T1 and ECV from the myocardium between the proposed method and MOLLI. The bull's-eye plot and the bar plot of the pre-contrast T1 values showed an overall good agreement between the two methods within the myocardium across different ROIs (Figure 9). The average pre-contrast myocardial T1 for 6 subjects is 1250 ms for the proposed method and 1253 ms for MOLLI. Similar results were shown for case of the estimated ECV values, except for the apical region which, we hypothesize, is partly caused by partial volume effect and/or banding artifacts in the MOLLI ECV maps (Figure 10). The average myocardial ECV for 6 subjects is 31.7 for the proposed method and 33.08 for MOLLI.

# 4. Discussion

In this work, we present a novel method that allows free-breathing 3D ECV mapping of the heart. The presented work displays two unique features. First, a recently developed manifold learning-based framework (49) is used for image reconstruction of the dynamic images of the heart. This framework leverages the intrinsic low-dimensional manifold structure of the underlying spatiotemporal signals to reconstruct high-framerate dynamic images that capture real time temporal variations due to respiratory motions and T1 contrast changes that are introduced by inversion recovery and/or post-injection contrast agent concentration variations. The LTSA model is a nonlinear generalization of the conventional linear models such as the low-rank or low-rank tensor model and can further reduce the number of unknowns in the resultant image reconstruction problem, leading to improved image quality. Results from this work show improved performance using this model compared to the low-rank and sparsity model in terms of the image reconstruction quality of the dynamic MR images (Figures 2 and 3) and the subsequent T1 estimation for both pre- and post-contrast cases (Supporting Information Figures S2 and S3) (54). Second, the real-time imaging capability of the LTSA model allows the use of a time-varying T1 model in post-contrast T1 mapping to account for the temporal changes in the contrast agent concentration. Results from this work indicate a relatively good approximation of this model, showing comparable results of the pre- and post-contrast T1, and the subsequent ECV estimations when compared to those from the MOLLI method (Figures 4-10).

Transmit $B_1$ ($B_1^+$) correction needs to be considered for accurate estimation of pre- and/or post-contrast T1 in the heart. In this work, joint $T_1/B_1^+$ mapping method (21) is utilized in consideration of the $B_1^+$ inhomogeneity of the heart at 3T. While the estimated cardiac $T_1$ maps were compared with the reference MOLLI method and showed good agreement, it is difficult to directly validate the estimated $B_1^+$ maps due to the lack of reference cardiac $B_1^+$ mapping method. To address this issue, the performance of the method was further validated through an in vivo brain study. Results show overall good agreement between the $T_1$ and $B_1^+$ estimations from the joint $T_1/B_1^+$ mapping method and those from the reference methods (Supporting Information Figures S4, S5 and S6). Supporting Information Figure S4 shows good agreement between the estimated $T_1$ from the joint $T_1/B_1^+$ mapping method and those from the reference inversion recovery with fast spin echo (IR-FSE) method within the gray and white matter regions of the brain. Good agreement between the estimated $B_1^+$ from the joint $T_1/B_1^+$ mapping method and those from the reference actual flip angle (AFI) method is shown, both qualitatively (Supporting Information Figure

S5) and quantitatively (Supporting Information Figure S6). Notably, the estimated T1 from the joint $T_1/B_1^+$ mapping method is similar for each tissue type, despite the spatial variation of the $B_1^+$.

$B_1^+$ correction for pre- and post-contrast $T_1$ estimation may not be necessary if ECV values are the only parameters of interest. We have conducted a preliminary investigation comparing ECV estimation with and without $B_1^+$ correction in the pre- and post-contrast $T_1$ estimations. The results are shown in Supporting Information Figure S9. Overall, the ECV estimation performance was comparable in terms of accuracy and precision between cases with and without $B_1^+$ correction in the pre- and post-contrast $T_1$ estimations (Supporting Information Figure S9). This may be because the estimated pre- and post-contrast T1 values of the myocardium and blood pool shared the same bias caused by B1+ inhomogeneity. Considering the small number of subjects in this study, further investigation is necessary for validation of this finding.

In vivo 3D post-contrast T1 mapping is challenging due to the time-varying T1 after the contrast agent injection. In this work, we administered the contrast agent with bolus injection and assumed that the post-contrast T1 varies linearly with time in vivo. However, this assumption may not hold well in certain circumstances, since literature suggests a linear relationship between relaxivity and time post-contrast injection in vivo when the contrast agent is administered with bolus injection (44,55). Supporting Information Figure S7 shows the variations in post-contrast T1 and ECV estimation with this assumption depending on the acquisition time of the data over different post-contrast injection times. Results show higher deviations from the expected post-contrast T1 values when the early frames after contrast agent injection were included for T1 estimation (Supporting Information Figure S7). Nonetheless, this may not be an issue for ECV estimation using data acquired with sufficient delay after post-contrast injection, since the ECV may be well approximated as constant with sufficient delay after post-contrast injection (i.e., when $\Delta R_{1,myocardium}/\Delta R_{1,blood}$, the ratio of the difference between pre- and post-contrast relaxivities of myocardium and blood, is maintained) (44,55). In support of this, preliminary investigation showed a relatively stable estimation of myocardial ECV over time with sufficient delay after post-contrast agent injection (Supporting Information Figure S7). Additionally, good agreement was observed between the ECV estimations from modeling the post-contrast T1 as static (i.e., constant over time) and dynamic (i.e., linearly increasing over time) (Supporting Information Figure S10). This implies that if ECVs are the only parameters of interest, they could be estimated while assuming static post-contrast T1 in a certain time window after contrast agent injection. Further investigation is necessary for validation of this finding.

Additional preliminary investigation was performed to show similar estimation performance of ECV between the cases of assuming linearly changing T1 or R1 with time, for data acquired with sufficient delay after post-contrast agent injection (data not shown). The similar performance between the cases of assuming linearly changing T1 or R1 with time may be due to the low speed of contrast wash-out, which makes a linear approximation sufficient in either domain. Alternatively, administration of contrast agent with bolus injection followed by continuous infusion may be employed to reach an equilibrium state for the T1 or R1 and to simplify the T1 estimation after contrast agent injection (44). However, the tradeoff between the complexity of imaging, subject comfort, and simplicity in T1 estimation needs to be carefully considered for this approach since introducing continuous infusion can create additional constraints during acquisition and burden on subjects. Further investigation is necessary to validate the feasibility and performance of this method in clinical practice.

This work has several limitations. First, the current work suffers from misregistration errors between pre- and post-contrast T1 maps. A non-rigid image registration method was used in this work to register between the pre- and post-contrast T1 maps, which is not ideal since the heart is an organ in which the size and shape change with cardiac and respiratory motions. Deformable image registration method may be used to further improve the results shown in this work. Second, the computation time to perform T1 fitting was relatively long for the post-contrast case in this work. The computation time may be further reduced by employing a higher-performance computer for computation or by reducing the stepsize and/or the range of the grid used for dictionary generation through optimizations. Third, data acquisition in this work was performed with an ECG-gated IR scheme and the results in this work may contain a similar bias as MOLLI due to blood flow and magnetization transfer effects. Lastly, only healthy volunteers were recruited for this work and no histological validations were performed. Further work with patients with cardiovascular diseases (e.g., myocardial fibrosis) along with histological validations through heart surgery or endomyocardial biopsy will allow us to further validate the accuracy and value of the developed technique in clinical settings.

# 5. Conclusions

The proposed method allowed free-breathing, 3D ECV mapping in a practically feasible imaging time. The estimated ECV values from the proposed method were comparable to those from the existing method.

# Acknowledgements

This work was supported in part by the National Institutes of Health (K01EB030045, P41EB022544, R01CA165221, R01EB033582, R01HL137230, and T32EB013180). We thank Nicole DaSilva, Marina MacDonald-Soccorso, and Alex MacDonagh for their help with research coordination and blood sample collection.

# Appendix

We have the following image reconstruction problem:

$$\arg\min_{\mathbf{T},\mathbf{L}} \frac{1}{2} \left\| \mathbf{\Omega}\left( \mathbf{F}_s \sum_{q=1}^Q \mathbf{TL}_q \mathbf{\Phi}_q^{\mathrm{T}} \mathbf{\Pi}_q \right) - d \right\|_2^2 + \frac{\mu_{\mathbf{T}}}{2} \| \mathbf{T} \|_F^2 + \frac{\mu_{\mathbf{L}}}{2} \| \mathbf{L} \|_F^2$$

$$+ \lambda_{\mathbf{T}} \| \mathcal{D}(\mathbf{TL}) \|_1 + \lambda_{\mathbf{L}} \| vec(\mathbf{L}) \|_1 \tag{A.1}$$

We propose a nested version of the ADMM algorithm to solve the optimization problem, yielding the following sub-problems:

$$\mathbf{T}^{(k+1)} = \arg\min_{\mathbf{T}} \frac{1}{2} \left\| \mathbf{\Omega}\left( \mathbf{F}_s \sum_{q=1}^Q \mathbf{TL}_q^{(k)} \mathbf{\Phi}_q^{\mathrm{T}} \mathbf{\Pi}_q \right) - d \right\|_2^2 + \frac{\mu_{\mathbf{T}}}{2} \| \mathbf{T} \|_F^2$$

$$+ \frac{\rho_{out}}{2} \left\| \mathcal{D}(\mathbf{TL}^{(k)}) - \mathbf{G}_{\mathbf{T}}^{(k)} + \boldsymbol{\eta}_{out}^{(k)} \right\|_F^2, \tag{A.2}$$

$$\mathbf{L}^{(k+1)} = \arg\min_{\mathbf{L}} \frac{1}{2} \left\| \mathbf{\Omega}\left( \mathbf{F}_s \sum_{q=1}^Q \mathbf{T}^{(k+1)} \mathbf{L}_q \mathbf{\Phi}_q^{\mathrm{T}} \mathbf{\Pi}_q \right) - d \right\|_2^2 + \lambda_{\mathbf{L}} \| vec(\mathbf{L}) \|_1 + \frac{\mu_{\mathbf{L}}}{2} \| \mathbf{L} \|_F^2$$

$$+ \frac{\rho_{out}}{2} \left\| \mathcal{D}(\mathbf{T}^{(k+1)}\mathbf{L}) - \mathbf{G}_{\mathbf{T}}^{(k)} + \boldsymbol{\eta}_{out}^{(k)} \right\|_F^2, \tag{A.3}$$

$$\mathbf{G}_{\mathbf{T}}^{(k+1)} = \mathcal{S}_{\lambda_{\mathbf{T}}/\rho_{out}}(\mathcal{D}(\mathbf{T}^{(k+1)}\mathbf{L}^{(k+1)}) + \boldsymbol{\eta}_{out}^{(k)}), \tag{A.4}$$

$$\boldsymbol{\eta}_{out}^{(k+1)} = \boldsymbol{\eta}_{out}^{(k)} + \mathcal{D}(\mathbf{T}^{(k+1)}\mathbf{L}^{(k+1)}) - \mathbf{G}_{\mathbf{T}}^{(k+1)}, \tag{A.5}$$

where $\mathbf{G_T}$ is the augmented Lagrangian split variable, $\boldsymbol{\eta}_{out}$ is the dual variable, $\rho_{out}$ is a scalar relaxation parameter, and $\mathcal{S}_{\lambda_\mathbf{T}/\rho_{out}}$ is a soft-thresholding operator with threshold $\lambda_\mathbf{T}/\rho_{out}$. The optimization problem in Eq. (A.2) is convex and can be solved using the conjugate gradient method.

The update of $\mathbf{L}$ in Eq. (A.3) is performed using a nested ADMM, resulting in the following updates:

$$\boldsymbol{L}^{(k+1,l+1)} = \underset{\mathbf{L}}{\arg\min} \frac{1}{2} \left\| \boldsymbol{\Omega} \left( \mathbf{F}_s \sum_{q=1}^{Q} \mathbf{T}^{(k+1)} \mathbf{L}_q \boldsymbol{\Phi}_q^\mathsf{T} \boldsymbol{\Pi}_q \right) - \boldsymbol{d} \right\|_2^2 + \frac{\mu_\mathbf{L}}{2} \| \mathbf{L} \|_F^2$$

$$+ \frac{\rho_{in}}{2} \left\| \boldsymbol{L} - \mathbf{G}_\mathbf{L}^{(k+1,l)} + \boldsymbol{\eta}_{in}^{(k+1,l)} \right\|_F^2 \tag{A.6}$$

$$+ \frac{\rho_{out}}{2} \left\| \mathcal{D}(\mathbf{T}^{(k+1)}\boldsymbol{L}) - \boldsymbol{G}_\mathbf{T}^{(k)} + \boldsymbol{\eta}_{out}^{(k)} \right\|_F^2$$

$$\mathbf{G}_\mathbf{L}^{(k+1,l+1)} = \mathcal{S}_{\lambda_L/\rho_{in}}(\mathbf{L}^{(k+1,l+1)} + \boldsymbol{\eta}_{in}^{(k+1,l)}), \tag{A.7}$$

$$\boldsymbol{\eta}_{in}^{(k+1,l+1)} = \boldsymbol{\eta}_{in}^{(k+1,l)} + \mathbf{L}^{(k+1,l+1)} - \mathbf{G}_\mathbf{L}^{(k+1,l+1)}, \tag{A.8}$$

The first sub-problem in Eq. (A.6) can be readily solved using a conjugate gradient, while the other updates have a closed-form solution.

The following table summarizes the whole reconstruction procedure:

---

**Algorithm 1:** Sparsity constrained LTSA model for dynamic MR image reconstruction.

---

**Input:** Undersampled (k, t)-space data $\boldsymbol{d}$.

**Output:** Dynamic MR image $\mathbf{X}$.

    **A.** **Preparation of the data**

        Set the reconstruction rank $D$.

        Perform initial LR reconstruction and bin the dynamic data into $Q$ neighborhoods.

        $\forall q \in [\![1; Q]\!]$, calculate temporal bases $\boldsymbol{\Phi}_q$ using SVD.

    **B.** **Reconstruction of T and L**

        Set hyperparameters $N_{out}$, $N_{in}$, $\mu_T$, $\mu_L$, $\lambda_T$, $\lambda_L$, $\rho_{out}$ and $\rho_{in}$.

        $\forall q \in [\![1; Q]\!]$, initialize $\mathbf{L}_q^{(1)}$, $\mathbf{G}_\mathbf{T}^{(1)}$, and $\boldsymbol{\eta}_{out}^{(1)}$

        **for** $k = 1$ to $N_{out} - 1$ **do**

            Solve **Eq. (A.2)** and get $\mathbf{T}^{(k+1)}$.

            **for** $l = 1$ to $N_{in} - 1$ **do**

---

Solve **Eq. (A.6)** and get $\mathbf{L}^{(k,l+1)}$.

Soft threshold $\mathbf{G}_{\mathbf{L}}^{(l+1)}$ as in **Eq. (A.7).**

Update the dual variable $\boldsymbol{\eta}_{in}^{(l+1)}$ as in **Eq. (A.8).**

**end for**

$\mathbf{L}^{(k)}=\mathbf{L}^{(k,N_{in})}$

Soft threshold $\mathbf{G}_{\mathbf{T}}^{(k+1)}$ as in **Eq. (A.4).**

Update the dual variable $\boldsymbol{\eta}_{out}^{(k+1)}$ as in **Eq. (A.5).**

**end for**

Form dynamic image $\mathbf{X}=\sum_{q=1}^{Q} \mathbf{T}^{(N_{out})}\mathbf{L}_q^{(N_{out})}\boldsymbol{\Phi}_q^{\mathrm{T}}\boldsymbol{\Pi}_q$

# Figure legends

**Figure 1.** Schematic diagram of data acquisition for 3D cardiac ECV mapping. a) Schematic diagram of data acquisition along with the experimental timeline. ECG-gated 3D cardiac T1 mapping is performed before and after Gadolinium injection for the estimation of pre-contrast T1, post-contrast T1, and ECV. In an N-(M) imaging protocol, imaging and training datasets are acquired over N cardiac cycles followed by signal recovery over M cardiac cycles after a non-selective inversion pulse. A spatially selective inversion pulse and a one-dimensional navigator are applied at a sagittal plane at the apex of the right hemidiaphragm to track respiratory motion. Scans with MOLLI are additionally performed for comparison. b) Illustration of training data acquisition. A pulse sequence diagram for acquiring three $k$-space lines along the $k_x$, $k_y$, and $k_z$ directions across the center of the $k$-space is shown with a schematic diagram displaying the $k$-space trajectory traversed by each readout gradients. c) Illustration of imaging data acquisition. A pulse sequence diagram for acquisition with randomized order of $k_z$ and spoke angles is shown with a schematic diagram displaying the traversed $k$-space trajectory.

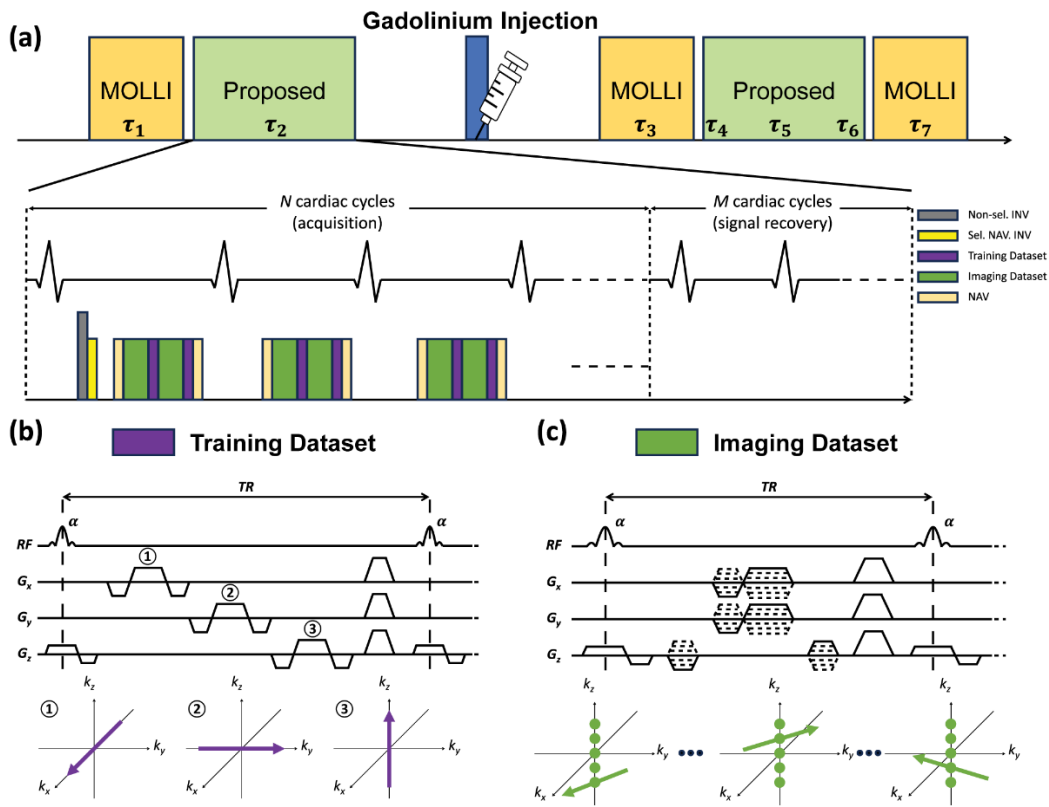

**Figure 2.** Comparison of reconstruction results between the LR-based and proposed LTSA-based method. Representative results from Subject 1 are shown for different frames.

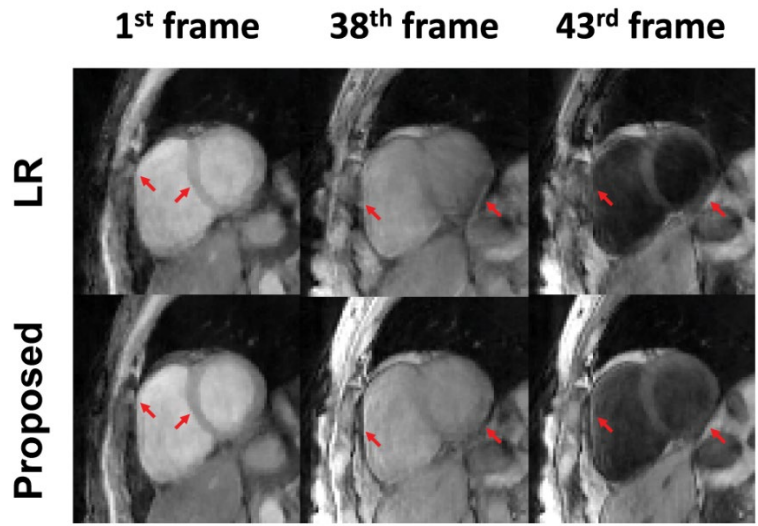

**Figure 3.** Results of image reconstruction from the proposed LTSA-based method. Representative results from Subject 1 are shown for different frames and slice indices. (a) Results from pre-contrast injection acquisition. Reconstructed images of different frames at a fixed slice index of 25 (top row) and different slice indices at a fixed frame index of 2 (bottom row) are shown. (b) Results from post-contrast injection acquisition. Reconstructed images of different frames at a fixed slice index of 25 (top row) and different slice indices at a fixed frame index of 4 (bottom row) are shown.

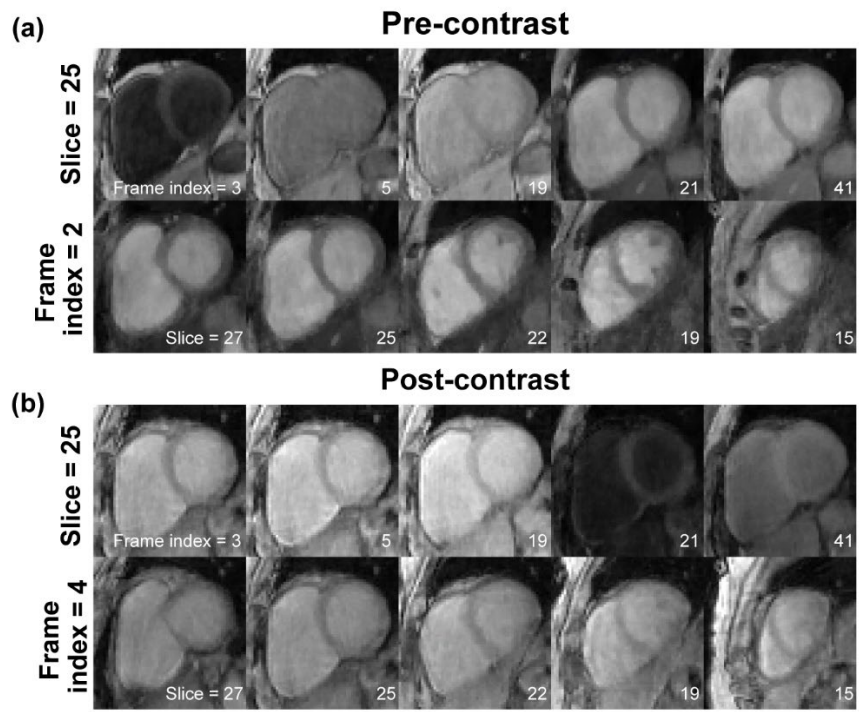

**Figure 4.** Pre-contrast $T_1$ mapping results. Representative results from Subject 3 are shown. (a) Pre-contrast 3D $T_1$ map from the proposed method evaluated at time $\tau_2$. $T_1$ maps in the short-axis view from the basal to apical regions of the heart are shown. (b) Comparison of pre-contrast $T_1$ maps between the proposed method and MOLLI. The $T_1$ maps from the proposed method at slice locations in the basal (red box), mid (blue box), and apical (green box) regions of the heart are compared to those from MOLLI acquired at time $\tau_1$.

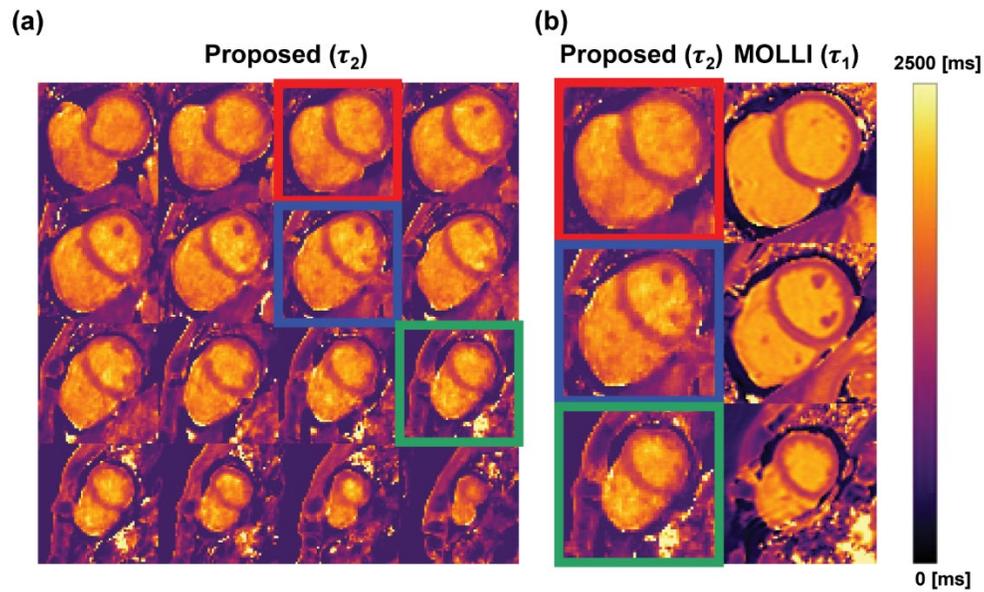

**Figure 5.** Post-contrast $T_1$ mapping results. Representative results from Subject 3 are shown. (a) Post-contrast 3D $T_1$ map from the proposed method evaluated at time $\tau_5$. $T_1$ maps in the short-axis view from the basal to apical regions of the heart are shown. (b) Comparison of post-contrast $T_1$ maps between the proposed method and MOLLI. The $T_1$ maps from the proposed method at slice locations in the basal (red box), mid (blue box), and apical (green box) regions of the heart are compared to those from MOLLI acquired at time $\tau_7$. The difference in $T_1$ values between the $T_1$ maps from the proposed method and MOLLI is presumed to be due to the difference in acquisition time after the administration of the contrast agent.

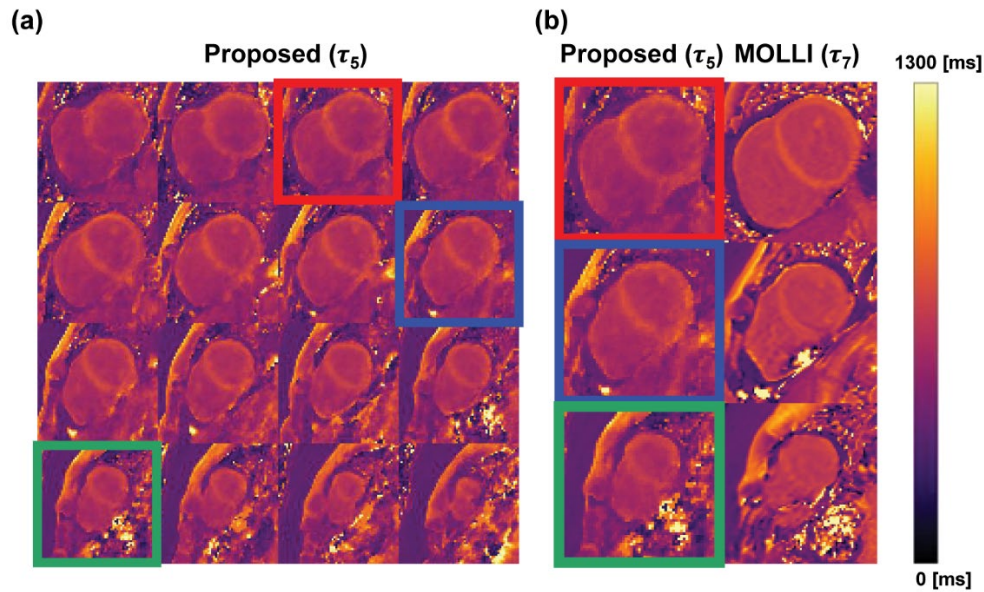

**Figure 6.** Evaluation of post-contrast $T_1$ estimated from the proposed method. Representative results from Subject 3 are shown. (a) Scatter plot of estimated post-contrast myocardial $T_1$ versus post-contrast injection time. The mean (black dot) and standard deviation (error bar) of the estimated post-contrast myocardial $T_1$ from the proposed method for times $\tau_4$, $\tau_5$, and $\tau_6$ are shown along with those from MOLLI acquired at times $\tau_3$ and $\tau_7$. (b) Post-contrast $T_1$ maps used for evaluation. The voxels within the myocardium from the basal and mid-slice locations were used for evaluation in (a). The apical slice region was excluded for analysis due to the banding artifacts caused by the bSSFP readout in the results obtained by MOLLI. Note the overall trend of increase in estimated myocardial $T_1$ value with post-contrast injection time.

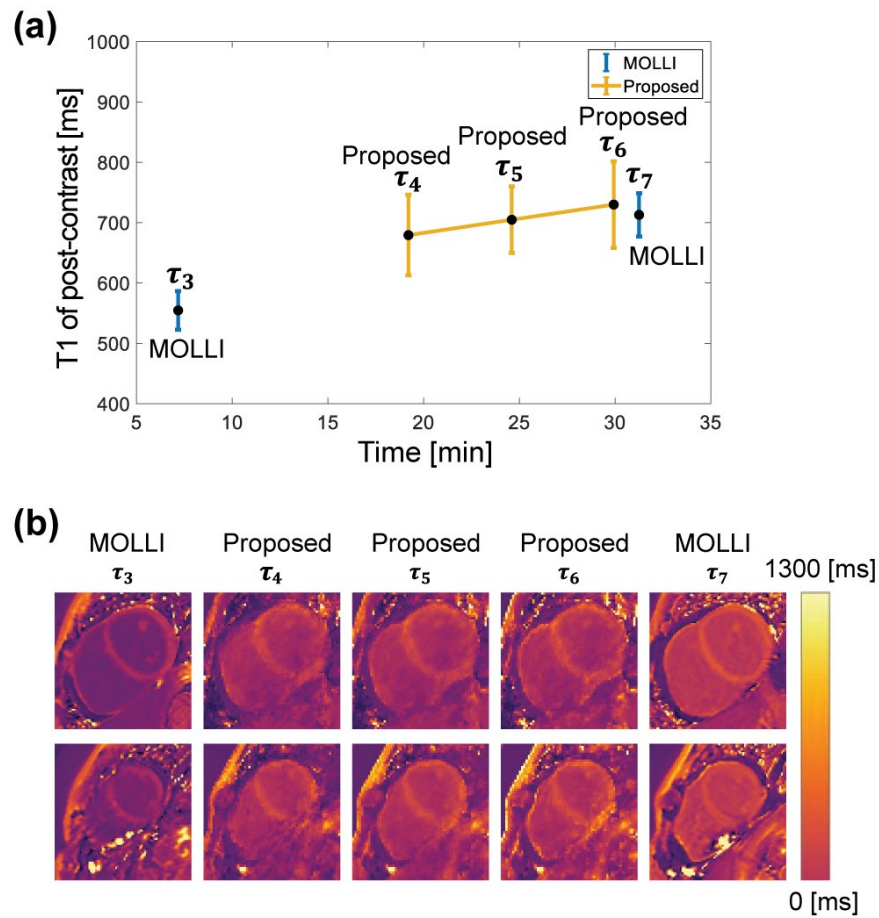

**Figure 7.** ECV mapping results. Representative results from Subject 3 are shown. (a) Three-dimensional ECV map from the proposed method using post-contrast $T_1$ maps evaluated at time $\tau_5$. ECV maps in the short-axis view from the basal to apical regions of the heart are shown. (b) Comparison of ECV maps between the proposed method and MOLLI. The ECV maps from the proposed method at slice locations in the basal (red box), mid (black box), and apical (green box) regions of the heart are compared to those from MOLLI using post-contrast $T_1$ maps acquired at time $\tau_7$.

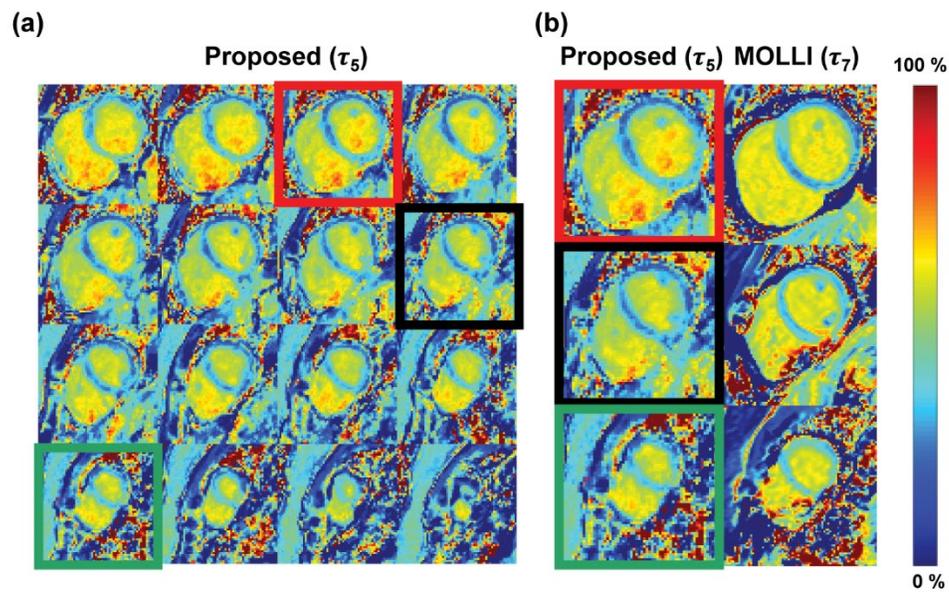

**Figure 8.** ECV mapping results from two additional subjects. Three-dimensional ECV maps estimated from subjects 2 and 5 using the proposed method are shown. The ECV maps from MOLLI corresponding to the same slice locations in the basal (red box), mid (black box), and apical (green box) regions of the heart are shown on the side for comparison.

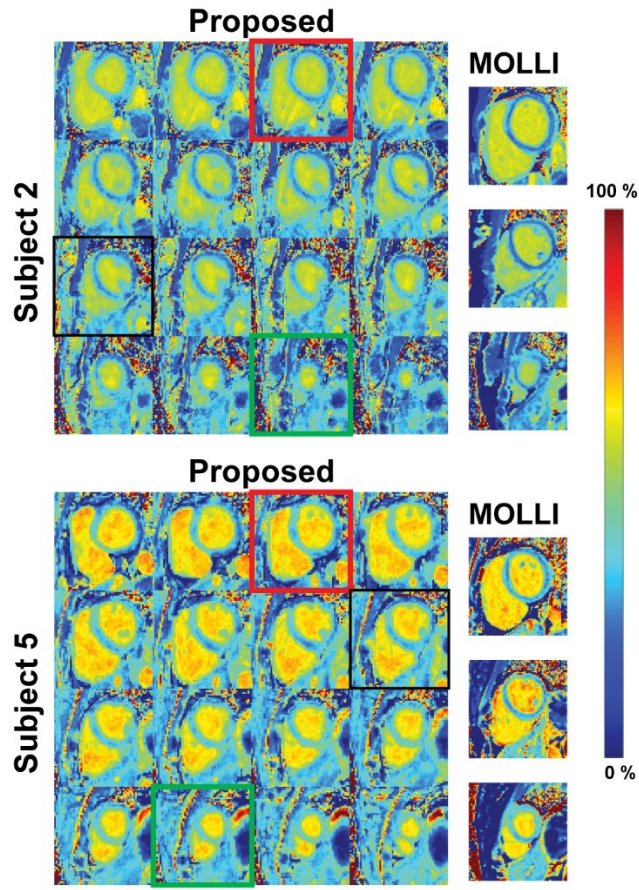

**Figure 9.** Comparison of estimated pre-contrast $T_1$ between the proposed method and MOLLI. (a) Bull's eye plot of mean myocardial $T_1$ estimated from the proposed method. (b) Bull's eye plot of mean myocardial $T_1$ estimated from MOLLI. (c) Bar plots of the estimated pre-contrast $T_1$ values for the basal anterior (BA), basal anteroseptal (BAS), basal inferoseptal (BIS), basal inferior (BI), basal inferolateral (BIL), basal anterolateral (BAL), midcavity anterior (MA), midcavity anteroseptal (MAS), midcavity inferoseptal (MIS), midcavity inferior (MI), midcavity inferolateral (MIL), midcavity anterolateral (MAL), apical anterior (AA), apical septal (AS), apical inferior (AI), and apical lateral (AL) regions of the left ventricular myocardium. The error bars denote the standard deviation of the mean across the subjects for each segment.

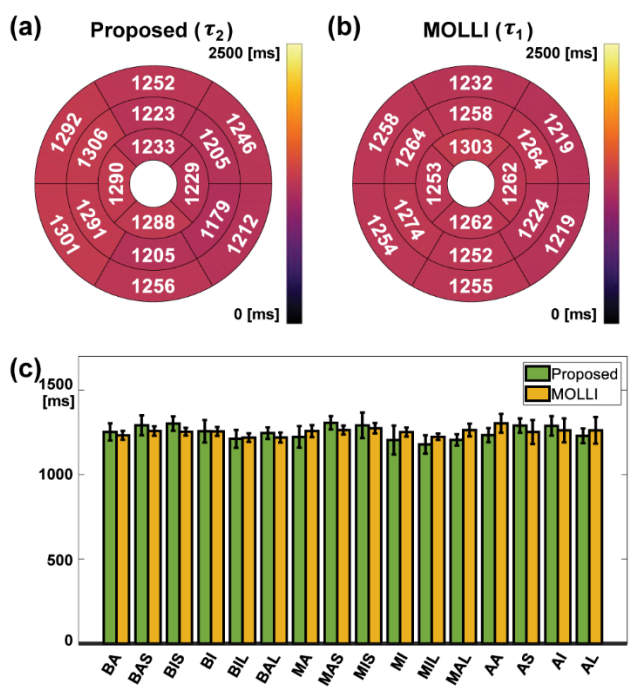

**Figure 10.** Comparison of estimated ECV between the proposed method and MOLLI. (a) Bull's eye plot of mean myocardial ECV estimated from the proposed method using post-contrast $T_1$ values evaluated at time $\tau_5$. (b) Bull's eye plot of mean myocardial ECV estimated from MOLLI with post-contrast acquisitions made at time $\tau_7$. (c) Bar plots of the estimated ECV values for the basal anterior (BA), basal anteroseptal (BAS), basal inferoseptal (BIS), basal inferior (BI), basal inferolateral (BIL), basal anterolateral (BAL), midcavity anterior (MA), midcavity anteroseptal (MAS), midcavity inferoseptal (MIS), midcavity inferior (MI), midcavity inferolateral (MIL), midcavity anterolateral (MAL), apical anterior (AA), apical septal (AS), apical inferior (AI), and apical lateral (AL) regions of the left ventricular myocardium. The error bars denote the standard deviation of the mean across the subjects for each segment.

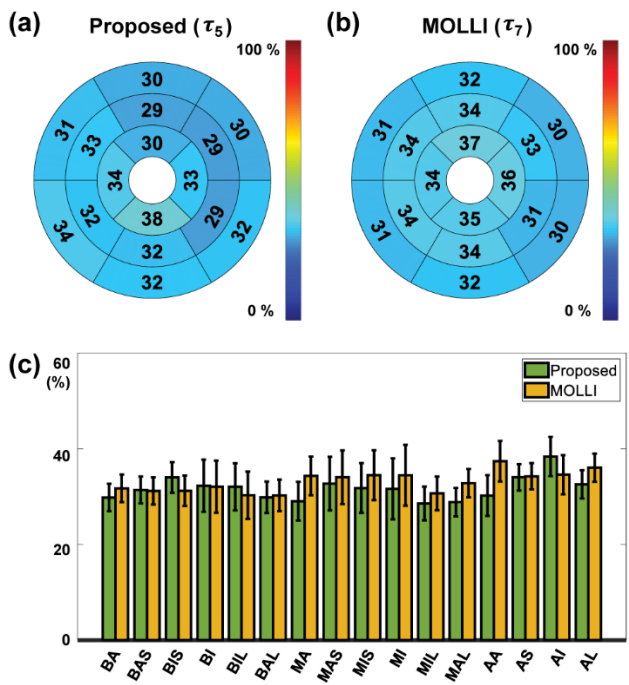

# Supporting materials

**Figure S1**. Image reconstruction results. (a) Reconstructed images using the LR method (1st column) and the LTSA method with (2nd column) and without (3rd column) the sparsity constraint on the matrix L. Reconstructed images at different time frames are shown. (b) Visual representation of the $L_q$ matrix of the 1st neighborhood for cases with and without the sparsity constraint on the matrix L. The $L_1$-norm value was calculated over the entire L matrix.

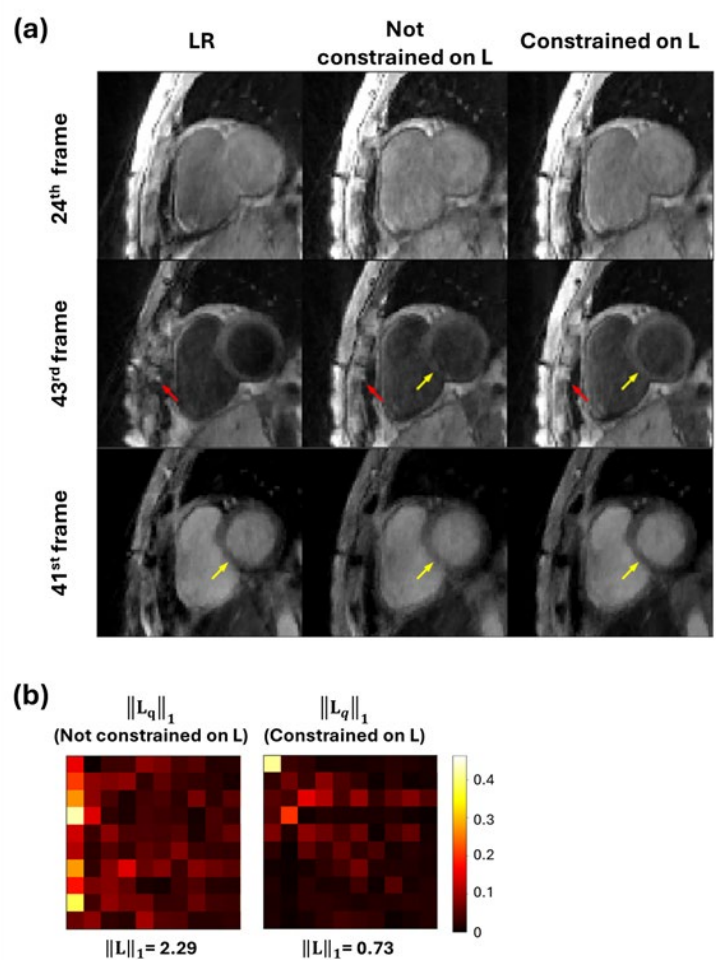

**Figure S2.** Comparison of pre-contrast $T_1$, post-contrast $T_1$, and ECV mapping results between the low-rank-based method, proposed LTSA-based method, and MOLLI. Results from Subject 1 are shown for slices in the mid and basal regions of the heart.

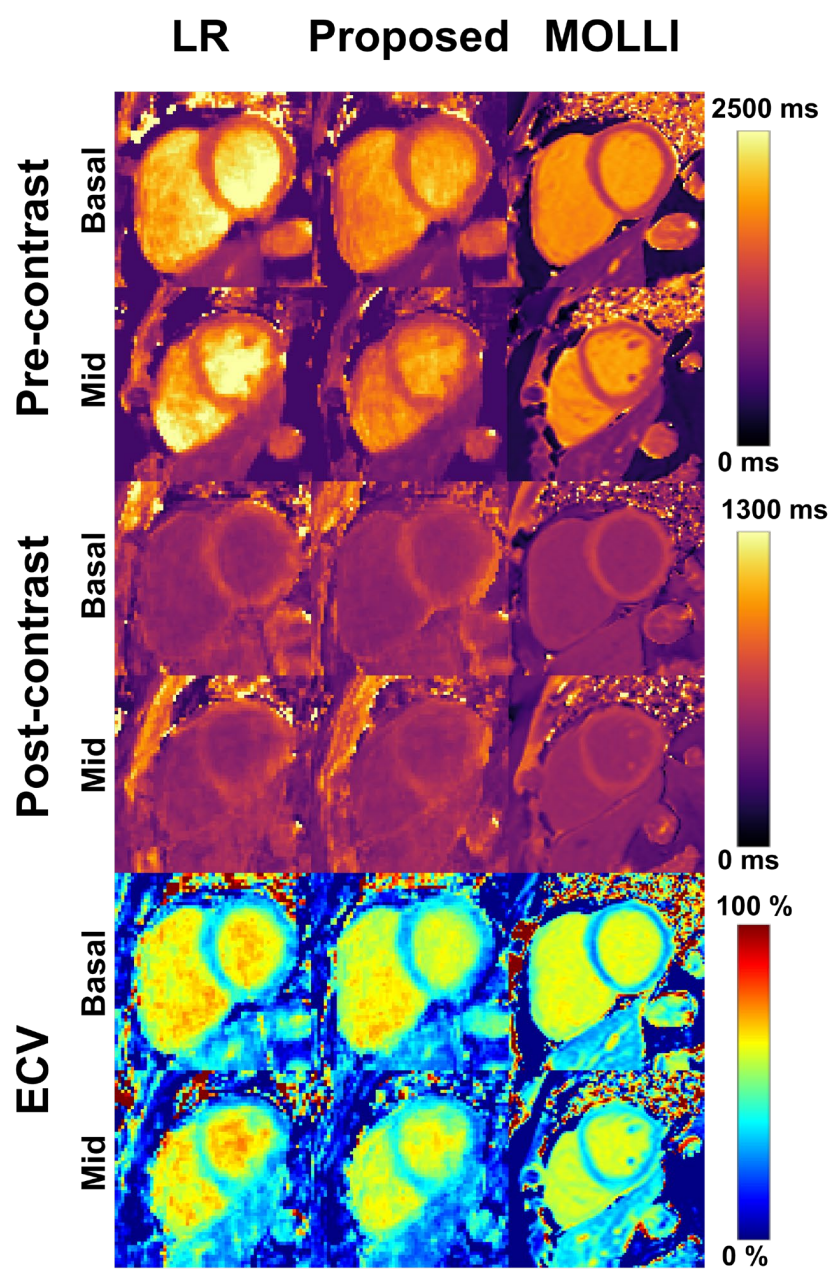

**Figure S3.** Comparison of $T_1$ and ECV estimation results between the low-rank-based method, proposed LTSA-based method, and MOLLI. The bar plots show the mean and standard deviation of pre- and post-contrast $T_1$ and ECV from a region-of-interest (ROI) within the blood or myocardium. The error bar represents the standard deviation within the ROI. The red horizontal line represents the mean $T_1$ or ECV from MOLLI. Results from Subject 1 are shown.

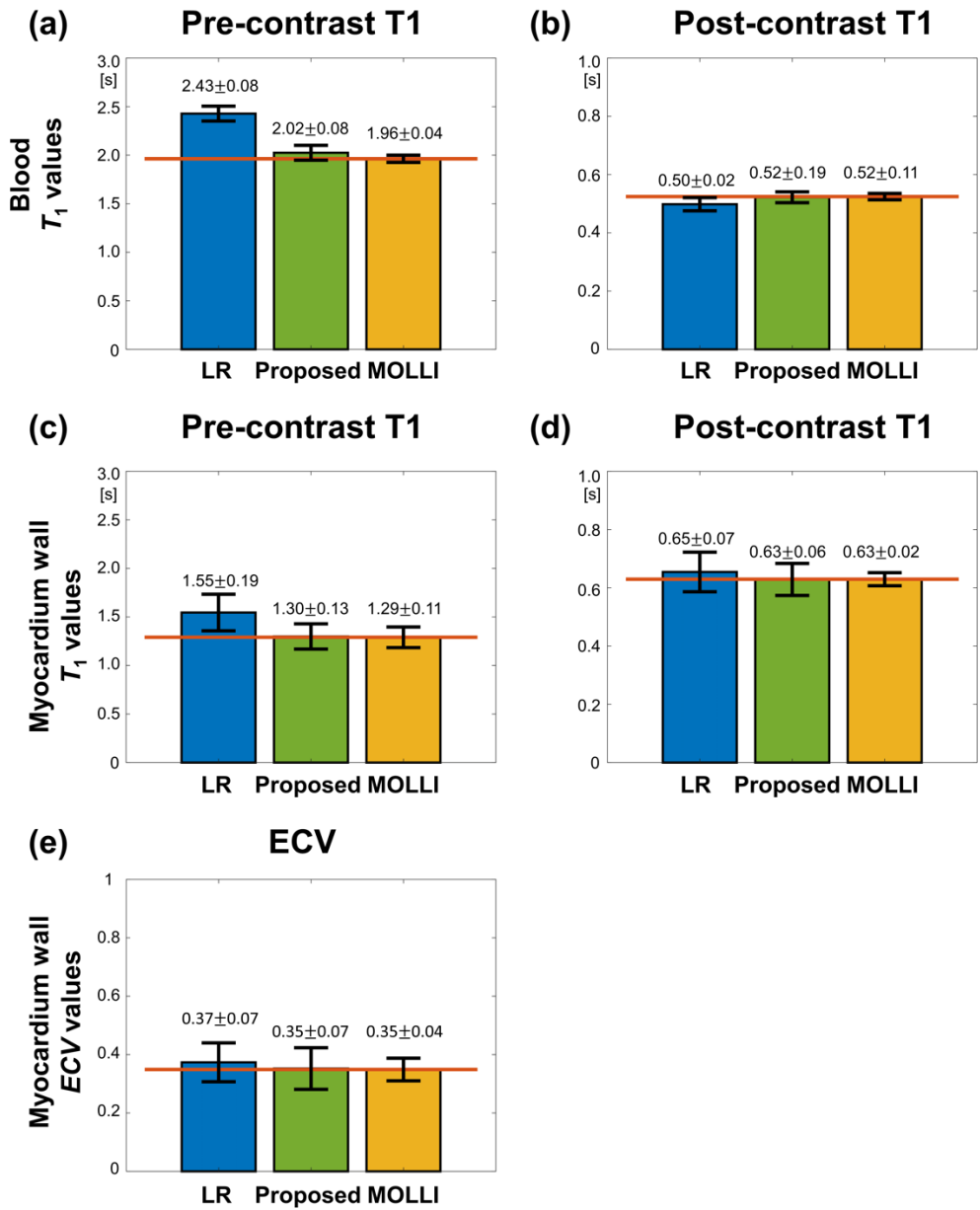

**Figure S4.** $T_1$ mapping results from an in vivo brain study. An in vivo brain imaging study was separately performed on a healthy volunteer to assess the joint $T_1$ and FA mapping results from the proposed method. For the proposed method, the same imaging sequence and parameters as in the cardiac studies were used with a simulated heart rate of 80 bpm, except the following imaging parameters: field-of-view (FOV) = 302×302×144 mm$^3$ and total acquisition time = 12.1 min. A reference 2D T1 map was obtained using an inversion recovery (IR) sequence with fast spin-echo (FSE) readout, with the following imaging parameters: FOV = 256×256 mm$^2$, matrix-size = 128×128, slice-thickness = 4.5 mm, TR = 10 s, echo spacing = 8.6, echo-train-length = 7, inversion time (TI) = 50/100/250/500/750/1000/1500/2000/2500/3000 ms, and total acquisition time = 34 min. (a) T1 map from the proposed method at the same slice location where 2D IR-FSE was performed. The estimated T1 within the gray and white matter regions were $1240 \pm 132$ ms and $869 \pm 30$ ms, respectively. (b) T1 map from the IR-FSE method. The estimated T1 within the gray and white matter regions were $1244 \pm 137$ ms and $864 \pm 34$ ms, respectively. (c) 3D T1 map from the proposed method.

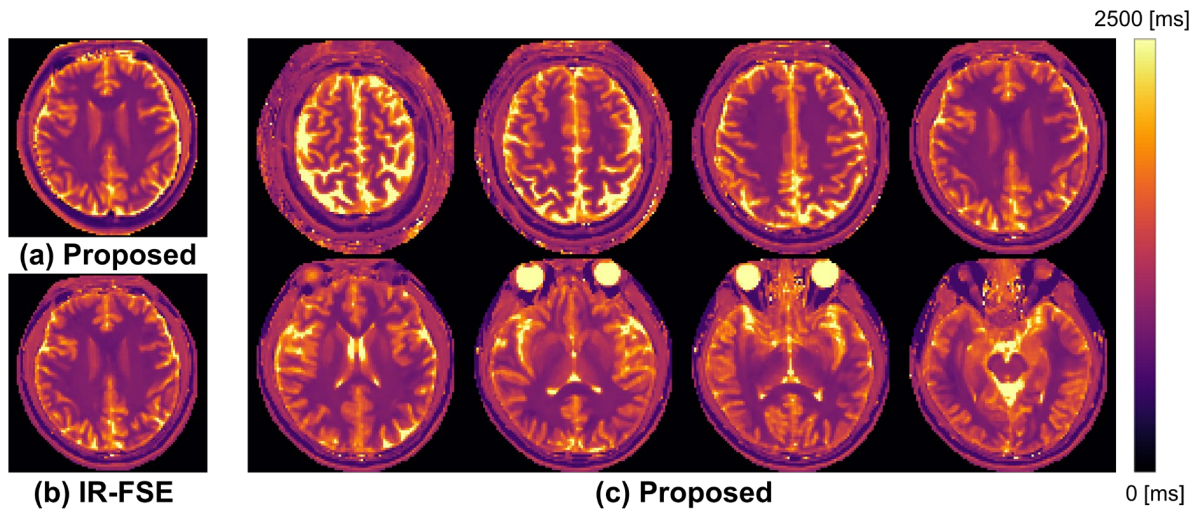

**Figure S5.** $B_1^+$ mapping results from an in vivo brain study. An in vivo brain imaging study was separately performed on a healthy volunteer to assess the joint $T_1$ and FA mapping results from the proposed method. The in vivo brain study was performed using the proposed method as described in Supporting Information Figure S3. Reference 3D $B_1^+$ map was obtained using actual flip-angle imaging (AFI) method with the following imaging parameters: FOV = 240×240×144 mm$^3$, matrix-size = 128×128×32, flip-angle = 60°, TR$_1$/TR$_2$ = 20/80 ms, TE = 2 ms, and total acquisition time = 10.7 min. (a) 3D $B_1^+$ map from the proposed method. (b) 3D $B_1^+$ map from the AFI method. Note the similarity in $B_1^+$ map between the proposed and AFI method.

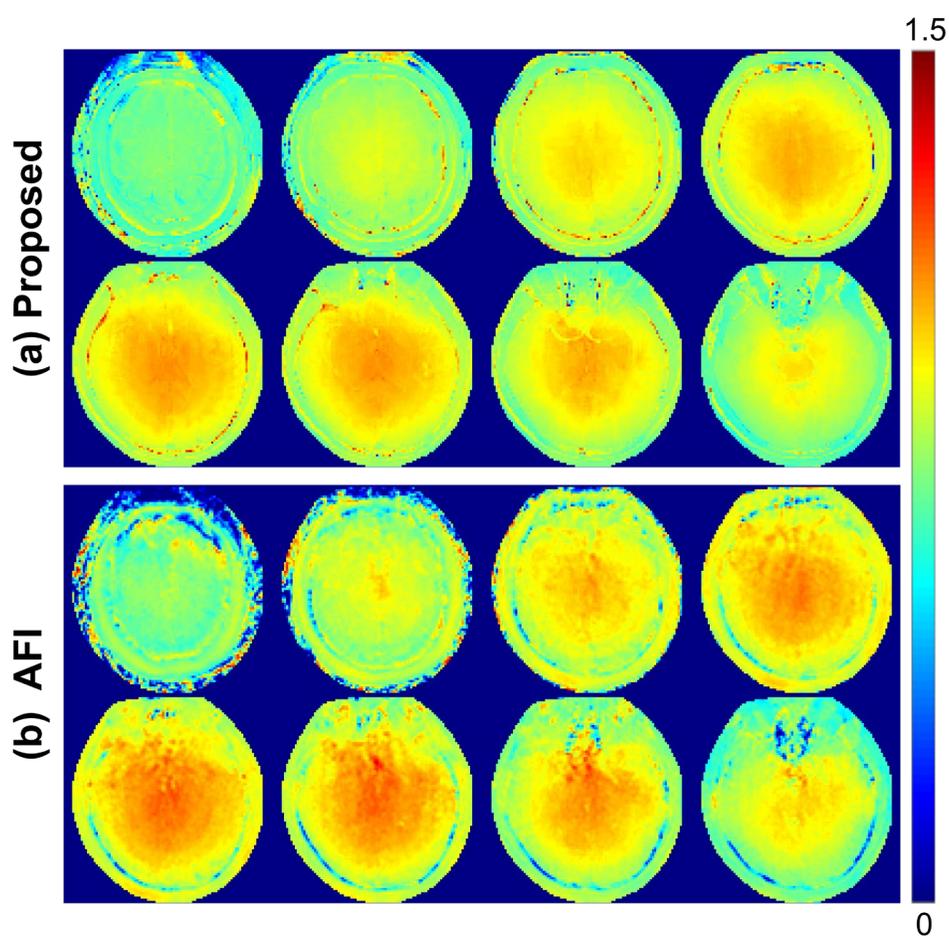

**Figure S6.** Comparison of estimated B$_1^+$ between the proposed and AFI method. (a) Scatter plot results showing voxel-wise comparison between the proposed and AFI method. The red line represents the line of identity and the blue line represents the line of regression. The results show high correlation and agreement between the two methods with $r^2 = 0.84$. (b) Bland-Altman plot showing voxel-wise comparison between the proposed and AFI method. The dotted blue lines represent the 95% confidence interval for the upper and lower limits of agreement. The red horizontal line marks the bias, 0.02.

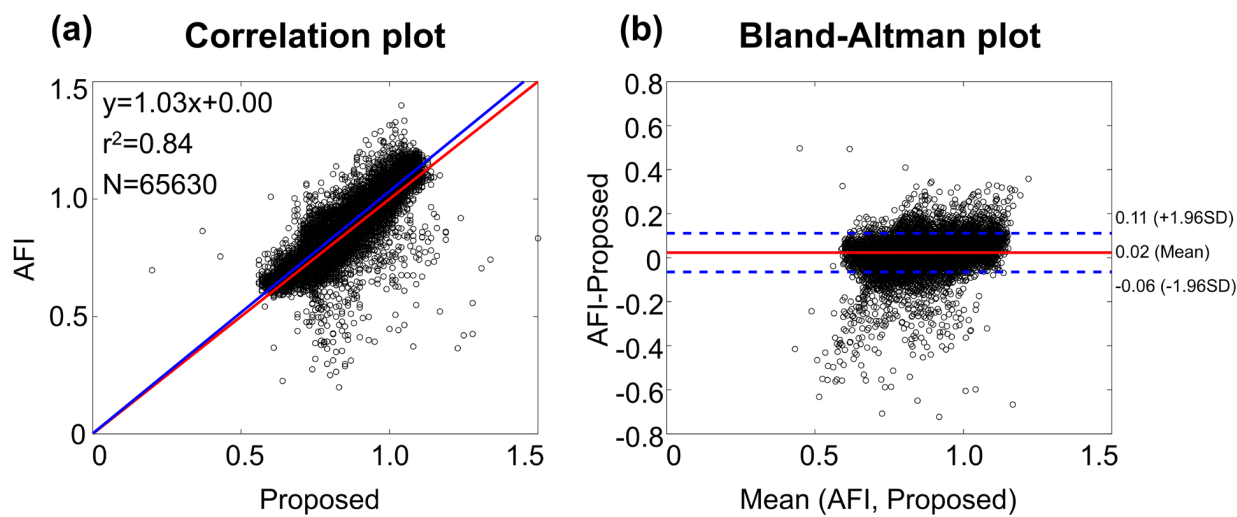

**Figure S7.** Analysis of post-contrast $T_1$ and ECV estimation over time after post-contrast agent injection. The mean and standard deviation of estimated post-contrast $T_1$ and ECV from a region-of-interest (ROI) within the myocardium are shown for Subjects 1 to 6, along with the schematic diagram of the experimental timeline. The mid-region of the heart without banding artifacts was used for assessment. The error bar represents the standard deviation within the ROI. Note the variations in post-contrast $T_1$ estimation depending on the acquisition time of the data acquired over different post-contrast injection times as shown in the case of Subject 3.

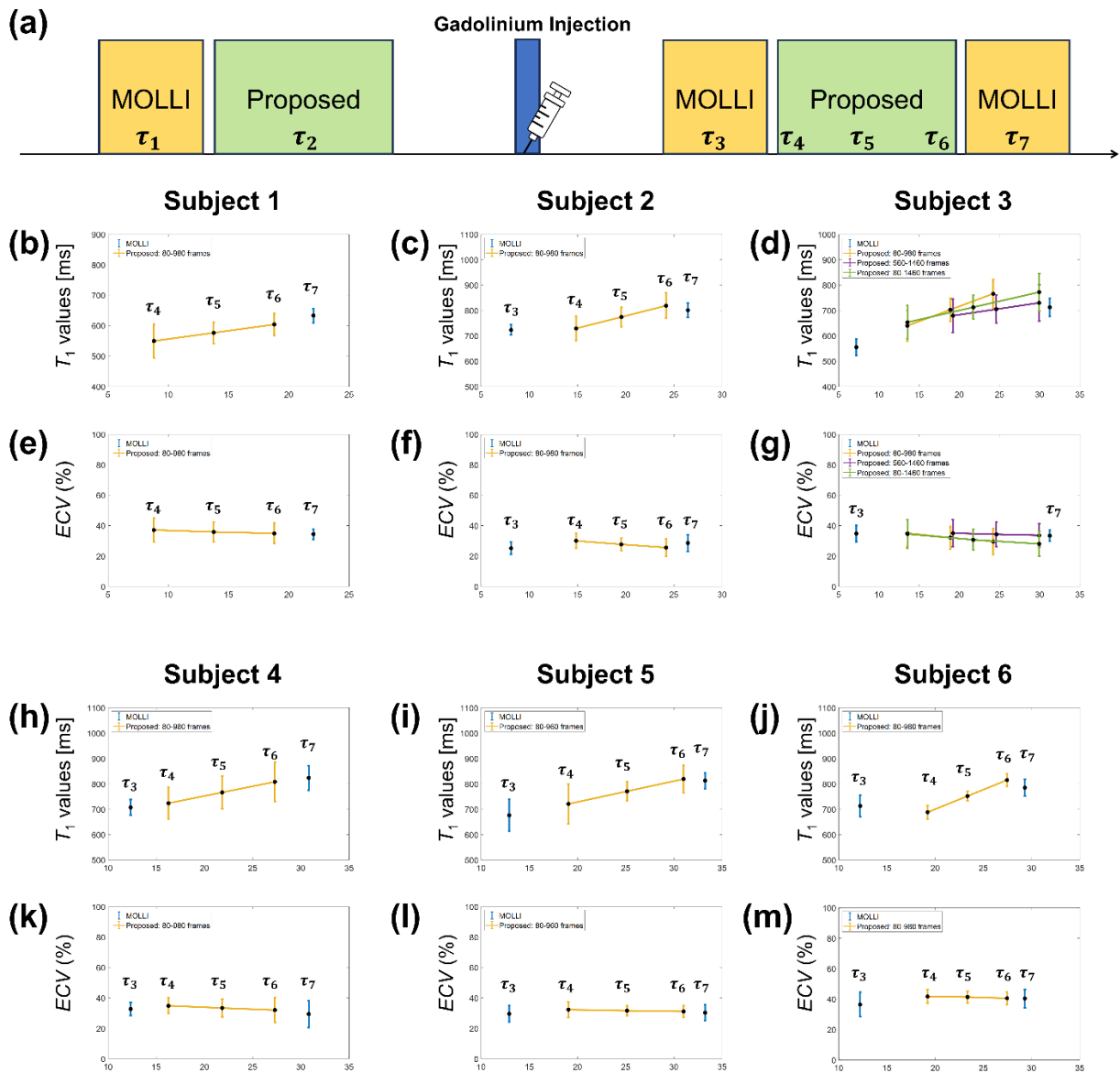

**Figure S8.** Comparison of pre-contrast $T_1$ and ECV between the proposed and MOLLI method. (a) Bland-Altman plot of ECV. (b) Bland-Altman plot of pre-contrast $T_1$. Segments of the myocardium according to the American Heart Association model without banding artifacts were used for assessment. The solid and dashed black lines denote the mean difference and the 95% confidence interval for the upper and lower limits of agreement, respectively.

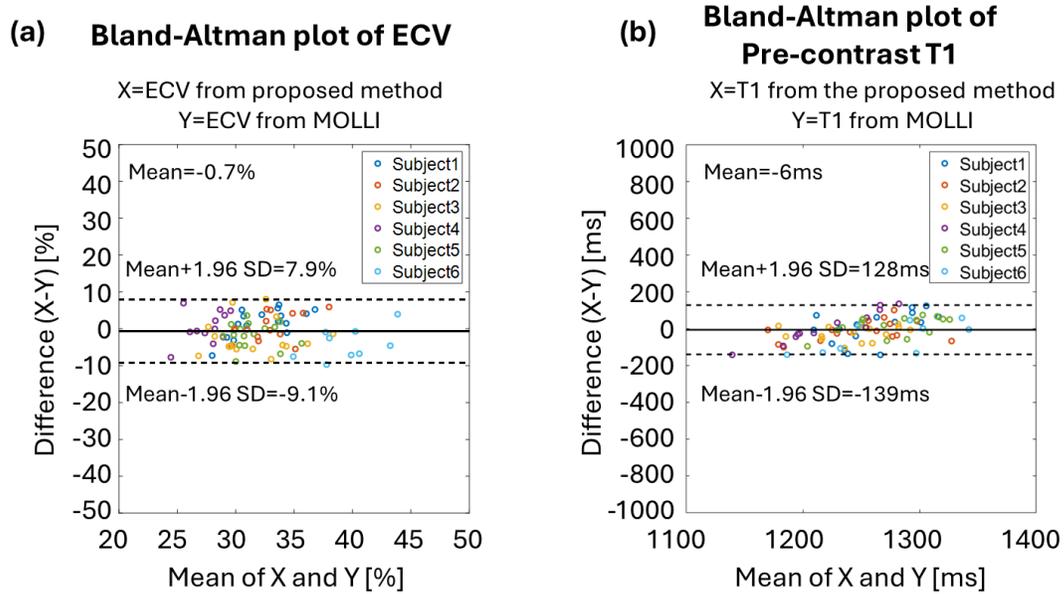

**Figure S9.** Comparison of ECV estimation with and without $B_1^+$ correction. (a) ECV maps from the proposed method without $B_1^+$ correction, proposed method with $B_1^+$ correction, and MOLLI. (b) Bland-Altman plots of ECV between the proposed method and MOLLI, with and without $B_1^+$ correction. Segments of the myocardium according to the American Heart Association model without banding artifacts were used for assessment. The solid and dashed black lines denote the mean difference and the 95% confidence interval for the upper and lower limits of agreement, respectively.

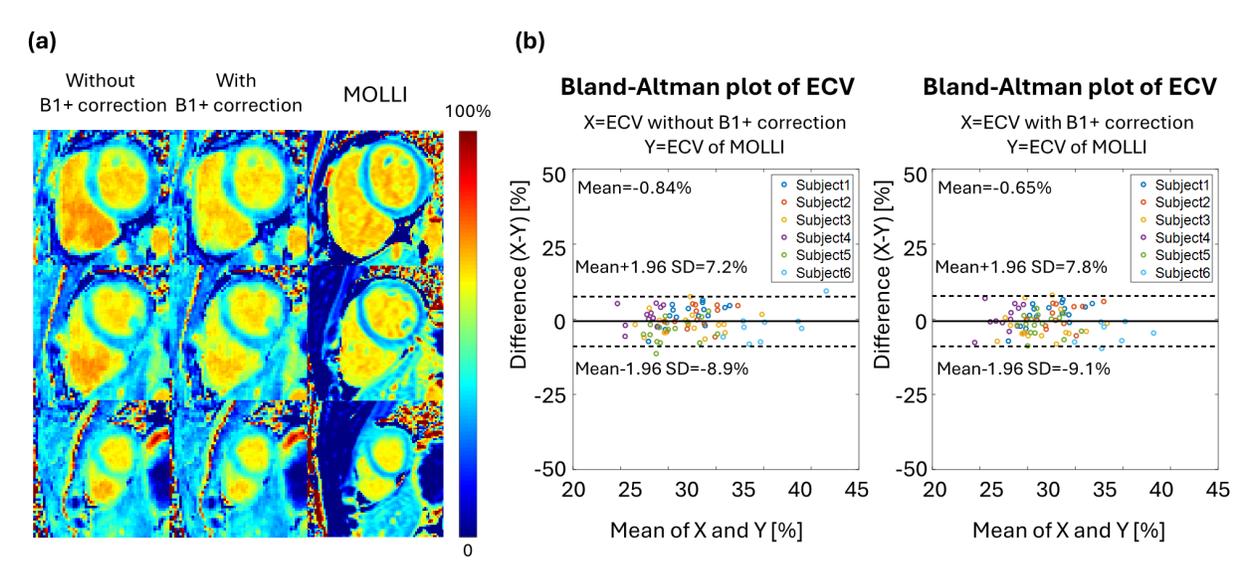

**Figure S10.** Analysis of static and dynamic $T_1$ modeling. (a) Post-contrast $T_1$ estimation from over time after post-contrast agent injection. The mean and standard deviation of estimated post-contrast $T_1$ from static and dynamic $T_1$ modeling and a region-of-interest (ROI) within the myocardium are shown for Subjects 1 to 6. The mid-region of the heart without banding artifacts was used for assessment. The error bar represents the standard deviation within the ROI. (b) Bland-Altman plot of ECV between the dynamic and static $T_1$ modeling. Segments of the myocardium according to the American Heart Association model without banding artifacts were used for assessment. The solid and dashed black lines denote the mean difference and the 95% confidence interval for the upper and lower limits of agreement, respectively.

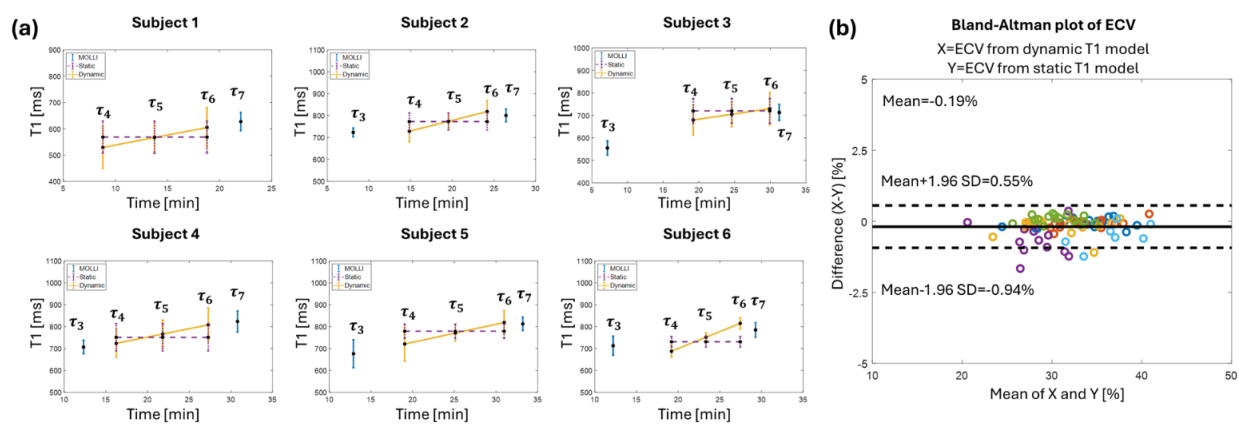